\documentclass[11pt,twoside]{article}
\usepackage{asp2010}

\resetcounters

\begin{document}

\title{Ideas for Advancing Code Sharing (A Different Kind of Hack Day)}
\author{Peter Teuben$^{1}$, Alice Allen$^2$, Bruce Berriman$^{3,4}$, Kimberly DuPrie$^2$, Robert J. Hanisch$^{4,5}$, Jessica Mink$^6$, Robert Nemiroff$^7$, Lior Shamir$^8$, Keith Shortridge$^9$, Mark Taylor$^{10}$,  and John Wallin$^{11}$
\affil{$^1$University of Maryland}
\affil{$^2$Astrophysics Source Code Library}
\affil{$^3$Infrared Processing and Analysis Center, California Institute of Technology}
\affil{$^4$Virtual Astronomical Observatory}
\affil{$^5$Space Telescope Science Institute}
\affil{$^6$Harvard-Smithsonian Center for Astrophysics}
\affil{$^7$Michigan Technological University}
\affil{$^8$Lawrence Technological University}
\affil{$^9$Australian Astronomical Observatory}
\affil{$^{10}$University of Bristol}
\affil{$^{11}$Middle Tennessee State University}}

\begin{abstract}

How do we as a community encourage the reuse of software for telescope
operations, data processing, and calibration? How can we support
making codes used in research available for others to examine?
Continuing the discussion from last year {\it Bring out your codes!}
BoF session, participants separated into groups to brainstorm ideas to 
mitigate factors which inhibit code sharing and nurture those which 
encourage code sharing. 
The BoF concluded with the sharing of ideas that arose from the
brainstorming sessions and a brief summary by the moderator.

\end{abstract}

\section{Introduction}

This Birds of a Feather (BoF) session was held to gather and discuss ideas on how 
to encourage the reuse of astronomical software, make computational research 
methods discoverable, remove barriers to code sharing, and better recognize and 
reward those who write software that enable science and enrich our community's 
efforts. This BoF builds directly on previous discussions \citep{AA_2012adass} and
presentations \citep{PT_2011adass}. Participants broke into groups for
two brainstorming sessions; the following questions were
prepared ahead of time to facilitate discussion:

\begin{enumerate}
{\em 

\item How do we encourage release even if the code is ``messy"?

\item How do we reduce expectations of support when software authors
  don't want to support software and still encourage code release?

\item How can universities be persuaded to change policies which
  prohibit software publication?

\item What can we do to encourage citations for codes?

\item Beyond citations, what can we do to give authors recognition for
  writing and releasing their software?

\item How can we measure the impact of a code on research and its
  value to the community?

\item What roles might journal publishers and funding agencies have in
  furthering code release, and how can the community influence them to
  take on that role?

\item What else can we do to have software release recognized as an
  essential part of research reproducibility?

}
\end{enumerate}

The question on recognition beyond citations (5) did not gather enough
interest and was dropped. Participants were also free to pose their own
questions; {\em what tools are available for sharing code?} was suggested
by Wil O'Mullane and discussed. Attendees were free to join any
group and to move to another question for the second brainstorming
period; a scribe in each group captured ideas. Throughout both
discussions, moderator Peter Teuben moved among the groups to follow
some of the brainstorming, and Nuria Lorente did the same and tweeted
out points being brought up in the conversations. She also monitored
the hashtags used for this session ({\tt \#adass2013} and {\tt
  \#asclnet}) to gather input from people not present who were
following the Twitter feed. After the brainstorming sessions, Teuben
moderated the presentation of the results and general discussion.

\section{Summary of Findings from the Group Discussions}

Convincing people to release software even if code is ``messy" (1) was the
most popular topic. 
Suggestions for mitigating this barrier to code release
included not allowing negative feedback on codes, offering a reward for
codes that are used even if they are messy, using GitHub to store and
exchange codes as a community practice to inspiring pieces of code,
just putting them out there, and running software as a web solution to
take pressure from the developer to fix code in a standalone pre-boxed
solution.

Those who discussed how to reduce possible expectations of support
when software authors release software they don't want to support (2)
suggested support be provided by the code's user community by using
forums such as {\tt stackoverflow} and {\tt astrobabel}.
One caveat to this was mentioned: that a code may not have enough users to make this a useful method
for its support. Someone opined that releasing a code through a public
repository such as GitHub is ``not really a release"; the term ``GitHub
mess" was used to describe such a release. This was countered by a
tweet stating ``that statement is completely flipped on its head by
stuff like {\tt @astropy}.  github IS THE support vehicle." Despite
the ``GitHub mess" comment, the group recognized that tools such as
GitHub and public repositories make it easier to enable that kind of
support for a code, thereby providing some relief to the author
of that expectation. Also mentioned by one of the two discussion
groups was whether the Astrophysics Source Code Library (ASCL)\footnote{www.ascl.net} should have metadata describing level of
support associated with entries, such as ``gold", ``bronze" and
``dirt".

%
In dealing with universities and code release (3), common practice is
to write and release code without regard to intellectual property
policies. A suggestion was to make sure code authors understand the
licenses and ramifications of each before talking with the university
attorneys; they should go into the discussion forearmed with
knowledge. Using NASA, NSF, or public funding requirements could also
be used as justification for release, as can pointing to existing
software released under a General Public License (GPL).  A desire 
for a class on licensing, perhaps at a future ADASS, was expressed.

The topic of recognition for authors who release codes focused primarily on
citations (4); participants suggested software authors include information
on how their codes should be cited right on the software's website,
and that citing the software's descriptive paper is almost always the
right thing to do. Citing ADASS publications, which may be the only
papers available for some codes, and standing up for one's work and
requesting it be cited were also discussed. It was pointed out no
standard practice exists, and perhaps a manifesto on software
citation, best practices, and policies should be developed
(e.g., \cite{Wilson2012}). Other 
suggestions included making sure one's code is reflected in ASCL and
ADS so it is citable and writing a paper in a journal that accepts
software and infrastructure.

Both groups who tackled measuring the impact and value of codes (6)
suggested the number of downloads could be used to determine a code's
value, though further observed that using the number of users or downloads as a
measure can be tricky, since one download may be shared with lots of
others. Repeated downloads by a particular user or users would
indicate that there is greater traction for the code, that the
download was more than a failed first attempt, and code still being
downloaded years after release also indicates the community values
it. Another suggestion was to try to determine the percentage of use
in the community by looking at the potential audience for the code and
ascertain the level of use within that potential audience; a small
audience for a code but with full saturation means the software has
great impact. Citations do not necessarily measure impact nor use, and
measuring the kind of and level of impact a code may have, which could
include scientific, social, breakthrough, or a break down of barriers
to get to the next level of productivity or use, is
difficult. Conducting a survey and rating codes were also suggested.

Discussion participants said funding agencies could clarify requirements and
policies (7); they further suggested funders should realize that
making code available is not free, and that if agencies require
software sharing, they should be prepared to pay for it. Sharing code
as an element of documentation of research is important; put bluntly,
``if your code is your science, then you are not publishing your
science unless you are publishing your code." Sharing code with a
guarantee of support to ensure reproducibility is asking too much. A
question was raised that if there were an absolute requirement on
publication of code which ran against organizational policy, would
federal funding policies force institutional policies to change?

Having a common place already established to store code may make it
easier for code authors to release their work, which may help with
having software recognized as an essential part of research
reproducibility (8). A desire for a site similar to GitHub but static,
where code can follow you wherever you go, was expressed. The lack of
good coding practices needs to be addressed, opined one group, before
a code can be recognized, and schooling grad students in recommended
software practices or requiring computer science courses may help with
this. Participants suggested coders should be encouraged to advertise
that their code is publicly available; a counterpoint was made that
some say their code is available, but it is difficult to get to. This was
expressed as ``Saying code is publicly
available but ignoring requests for it, putting it somewhere obscure,
etc., is NOT code sharing." Another suggestion was to make licensing
requirements more transparent: if one works for an institute, it can
be quite difficult to figure out what licensing requirements are.

GitHub was suggested as a primary tool for sharing software, though
one participant in these discussions uses SourceForge for his
codes. Having an astronomy area with endorsements would be helpful, as
would more education on software practices, such as that provided by
Software Carpentry.

Documents provided by the scribes for the brainstorming sessions that 
capture the discussions more completely are
available online.\footnote{http://asterisk.apod.com/wp/?p=543}

\section{Conclusions}

In the familiar list of inhibiting factors, no major shifts in solutions
were found. Several times issues around better educating our community 
came up, through forums, as well as a paper on recommended software 
practices. It is clear to these authors that the ASCL can play a role in 
endorsing and contributing to software sharing in the community. 


\acknowledgements

Our thanks to Omar Laurino, Kai Polsterer, and Wil O'Mullane for
scribing, and to Nuria Lorente for tweeting discussion points,
comments, and questions throughout the BoF and serving as the voice
for those not present who provided input to the discussion.

\bibliography{B02}

\end{document}